\documentstyle[preprint,tighten,prd,aps,epsfig,eqsecnum]{revtex}

\def\bE{{\bf E}}
\def\bH{{\bf H}}
\def\bj{{\bf j}}
\def\bA{{\bf A}}
\def\bB{{\bf B}}
\def\bz{{\bf z}}

\def\bth{{\boldmath \theta}}
\def\bv{{\bf v}}
\begin{document}

\title{Flux tube dynamics in the dual superconductor}
\author{Melissa A. Lampert\thanks{Electronic 
        mail: melissa@albert.tau.ac.il}
        and Benjamin Svetitsky\thanks{Electronic 
        mail: bqs@julian.tau.ac.il}}
\address{School of Physics and Astronomy, Raymond and Beverly Sackler
Faculty of Exact Sciences, \\
Tel Aviv University, 69978 Tel Aviv, Israel}

\date{\today}
\maketitle
\begin{abstract}
We study plasma oscillations in a flux tube of the dual superconductor
model of 't~Hooft and Mandelstam.
A magnetic condensate is coupled to an electromagnetic field by its
dual vector potential, and fixed electric charges set up a flux
tube.
An electrically charged fluid (a quark plasma)
flows in the tube and screens the fixed
charges via plasma oscillations.
We investigate both Type I and Type II superconductors, with plasma
frequencies both above and below the threshold for radiation into the
Higgs vacuum.
We find strong radiation of electric flux into the superconductor
in all regimes, and argue that this invalidates the use of the simplest dual
superconductor model for dynamical problems.
\end{abstract}

%
\section{Introduction}
\label{sec:intro}

The confinement of color in quantum chromodynamics is explained if
color electric fields form flux tubes.  These flux tubes appear readily
in the bag model \cite{ref:bag}, but this is a geometric picture
rather than a dynamical one.  't~Hooft \cite{ref:thooft} and
Mandelstam \cite{ref:mandelstam} proposed that, just as magnetic flux
tubes form in a superconductor \cite{ref:abrikosov}, a condensation of
magnetic charge in the QCD vacuum would lead to the formation of
electric flux tubes and confinement.  This is the dual superconductor
picture of confinement.

The analytical study of the confining flux tube depends largely on a
classical, Abelian model.  One can connect this model to QCD with
't~Hooft's idea of Abelian dominance \cite{ref:oblique}.  Starting in
QCD, one fixes an Abelian gauge and asserts that the dominant degrees
of freedom are Abelian gauge fields (from a Cartan subalgebra) and
Abelian magnetic monopoles, with weak coupling to the non-Abelian
gauge fields which assume the guise of self-coupled charged fields.
The assumption that the effective interaction of the monopoles causes
their condensation leads immediately to the dual superconductor
picture.
The validity of this scenario is a subject of research and debate in
the lattice gauge theory community \cite{lattice_Abdom}.

In superconductivity, one studies the static structure of magnetic
flux tubes via a Landau-Ginzburg theory \cite{ref:lanlif}.  The
simplest Landau-Ginzburg Hamiltonian is that of the Abelian Higgs
model in three dimensions.  Classical solutions of the Abelian Higgs
theory have been considered as models of QCD's electric flux tube as
well \cite{ref:jands,patkos,alcock,ref:nair,ref:ball}.  
Suganuma, Sasaki, and Toki and their collaborators 
\cite{ref:DGL,ref:otherDGL},
using the formalism proposed by Suzuki \cite{ref:Suzuki}, have fixed the
effective coupling constants of the Abelian Higgs model by comparing its
flux tube with phenomenology.
There have also been attempts to do so by comparison with the flux tube that
emerges in lattice gauge theory \cite{ref:Wuppertal}.
Success in this program will establish a
Landau-Ginzburg effective Hamiltonian of QCD.

In this paper we present a study not of the statics of the confining
flux tube, but of its dynamics.  The flux tube of QCD is not a static
object.  Creation of flux tubes and their subsequent decay through
pair creation offer a detailed model of particle creation in $e^+e^-$,
$pp$, and $pA$ collisions \cite{ref:casher,ref:lund}.  In the case of
nucleus--nucleus collisions, the flux tube has an appreciable
transverse extent, so that it can be called a ``color capacitor'' if
the color field is coherent across it, or a ``color rope''
\cite{ref:biro} if it is not.  $q\bar q$ pairs and gluons are created
through the Schwinger pair creation process
\cite{ref:schwinger}, and they screen the field
in the flux tube while carrying away the energy in the form of
hadrons.  If the density of created particles is large enough, the
quarks and gluons form a quark--gluon plasma before the final
hadronization takes place.

Field-theoretic analysis of pair creation and back-reaction in the
flux tube \cite{ref:CEKMS} has shown the buildup of particle density
and subsequent plasma oscillations.  The methods were applied first to
a field region of infinite spatial extent, rather than to a finite
flux tube.  Eisenberg \cite{ref:jme} carried out a calculation in a
cylindrical flux tube of fixed radius, as suggested by the bag
model.\footnote{This calculation imposed superconducting boundary
conditions at the tube surface, instead of the more appropriate dual
superconductor.}  In our study here, the flux tube is a dynamical
field configuration, not a static geometric object.

Following \cite{ref:Suzuki} and \cite{ref:DGL}, 
we set up the electric flux tube in
classical electrodynamics coupled to an Abelian Higgs field via the
dual gauge field.  Thus the Higgs field represents a condensate of
magnetic monopoles; the dual Meissner effect confines {\em electric\/}
flux to flux tubes.  Starting from a static flux tube configuration,
we release a density of electric charges in the system and allow them
to accelerate and screen the electric field.  The weakening of the
electric field allows the flux tube to collapse, but the inertia of
the charges carries them into plasma oscillations that build up the
field strength again and force open the tube.  The coupling of the
plasma oscillations to the Higgs field making up the flux tube is the
main new feature in our work.

In Section~\ref{sec:eqns} we describe the dual superconductor theory.
The model contains an Abelian gauge field governed by Maxwell's
equations, with coupling to electric charges and currents on the one
hand and to a magnetic Higgs field on the other.  The electric
currents are treated hydrodynamically; the Higgs field is a classical
field that confines electric flux to the tube.  The coupling to the
Higgs field is accomplished by means of a dual vector potential, as
introduced by Zwanziger \cite{ref:zwanziger} and used subsequently in
\cite{ref:jands} for the static problem and in \cite{ref:DGL,ref:otherDGL}
for phenomenological modeling.  We impose cylindrical
symmetry and we assume $z$-independence along the flux tube in the
region far from its ends.  Nevertheless, we are forced to pay some
attention to what happens near the ends of the tube in order to define
potentials unambiguously.

We present numerical results in Section~\ref{sec:results}.  We set the
parameters of the theory in both the Type~I and Type~II regimes, and
show how plasma oscillations arise. We consider plasma frequencies
both above and below the cutoff for radiation into the Higgs vacuum.
The plasma oscillations are accompanied by changes in the radius of
the flux tube as the electric pressure from within decreases through
screening, only to increase again as the currents overshoot.

We believe that this paper is the first to address the dynamics of an
electric flux tube within this model.\footnote{Loh {\em et al.} \cite{ref:loh}
have simulated breaking of the flux tube in the Friedberg--Lee model
\cite{ref:Friedberg} which posits a confinement mechanism unrelated to
monopole condensation.}
It is amusing to
note that this physical situation has no counterpart in
superconductivity.  While our magnetic Higgs field appears in the
usual Landau-Ginzburg theory as an electrically charged condensate,
our electric currents cannot appear there because there are no
magnetic monopoles in nature.  An Abrikosov flux tube has to run all
the way to the boundary of the sample, where it joins onto the
external magnetic field.  Our electric flux tube, on the other hand,
has a finite (but large) length, and the charges at its ends can be
screened by the electric currents that flow in it.  We review in the
appendix the well-known phenomenon of flux quantization and why it has
no effect on a flux tube of finite length.
%
%
\section{Field equations in cylindrical geometry}
\label{sec:eqns}

Our classical model for the flux tube is electrodynamics coupled to a
scalar magnetic monopole field. The equations of motion are
Maxwell's equations with both electric and magnetic sources, plus the
Klein-Gordon equation for the monopoles.  The latter is coupled not to
the usual vector potential $A_{\mu}$ but to the dual potential
\cite{ref:zwanziger} $B_{\mu}$, and contains a self-interaction that
puts it into the Higgs phase.

\begin{figure}
   \begin{center} 
   \mbox{\epsfig{file=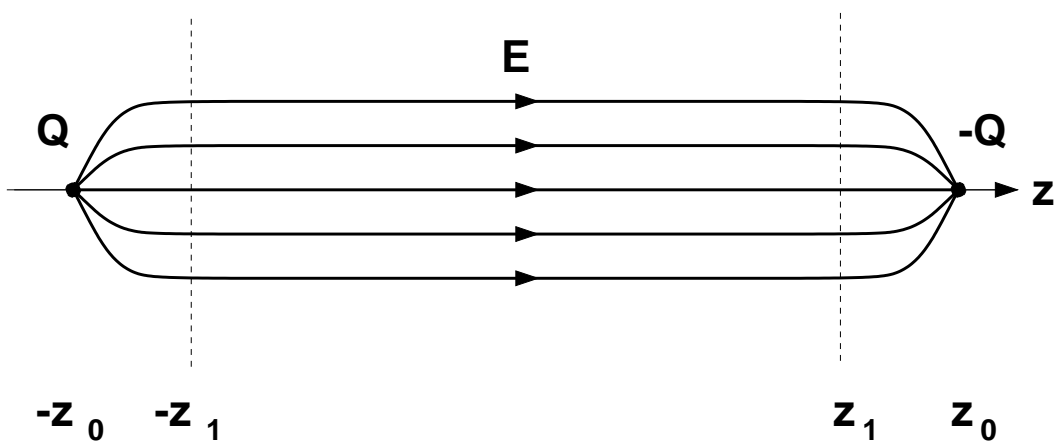,width=5.5in}}
   \caption{Geometry of the flux tube.}
   \label{fig:geometry}
   \end{center}
\end{figure}

We make a cylindrically symmetric {\em ansatz\/} for the fields, and 
furthermore assume
$z$-independence far from the sources at the ends of the tube.  We put
the electrically charged sources on the $z$ axis at $z=\pm z_0$, and
we assume that all the space-charge effects are localized there.  
The central region is defined by $|z|<z_1$, where
$z_1<z_0$ is chosen to exclude the end regions (see Fig.~\ref{fig:geometry}).

%
\subsection{Maxwell's equations}

Maxwell's equations with electric and magnetic currents are 
\begin{eqnarray}
   \partial_\mu F^{\mu\nu} &=& j_e^\nu \label{egauss4}\\
   \partial_\mu \tilde F^{\mu\nu} &=& j_g^\nu \label{mgauss4}
\>,
\end{eqnarray}
where $\tilde F^{\mu\nu} = \frac12 \epsilon^{\mu\nu\lambda\sigma}
F_{\lambda\sigma}$.  Eq.~(\ref{mgauss4}) replaces the Bianchi identity
of ordinary electrodynamics.  We can solve it with a vector potential,
but there is a new term,
\begin{equation}
   F^{\mu\nu} = \partial^\mu A^\nu - \partial^\nu A^\mu 
      + \epsilon^{\mu\nu\lambda\sigma} G_{\lambda\sigma}\>.
\label{direct1}
\end{equation}
Defining an arbitrary vector $n^\mu$, one finds that (\ref{mgauss4}) is
solved if
\begin{equation}
   G^{\mu\nu} = - n^\mu (n \cdot \partial)^{-1} j_g^\nu
\>,
\label{direct2}
\end{equation}
where $(n \cdot \partial)^{-1}$ represents an integration from
infinity with suitable boundary conditions.  The field equation for
$A_\mu$ then follows from (\ref{egauss4}).  Solving Maxwell's
equations in the opposite order, the same field strength can be
represented alternatively by a {\em dual\/} vector potential,
according to
\begin{equation}
   \tilde F^{\mu\nu} = \partial^\mu B^\nu - \partial^\nu B^\mu  
      + \epsilon^{\mu\nu\lambda\sigma} M_{\lambda\sigma}\>,
\label{dual1}
\end{equation}
with
\begin{equation}
   M^{\mu\nu} = - n^\mu (n \cdot \partial)^{-1} j_e^\nu
\>,
\label{dual2}
\end{equation}
in order to satisfy (\ref{egauss4}).  Now the field equation for
$B_\mu$ follows from (\ref{mgauss4}).  We will return to the vector
potentials below.

In three-dimensional notation,
\begin{mathletters}
   \begin{equation}
      \nabla \cdot \bE = \rho_e
   \end{equation}
   \begin{equation}
      \nabla \times \bH - \frac{\partial\bE}{\partial t} = \bj_e
   \label{eq:ampere}
   \end{equation}
   \begin{equation}
      \nabla \cdot \bH = \rho_g
   \end{equation}
   \begin{equation}
      \nabla \times \bE + \frac{\partial\bH}{\partial t} = -\bj_g
   \label{eq:faraday}
   \end{equation}
\end{mathletters}
In the aforementioned central region $|z|<z_1$,
we make the {\em ansatz}
\begin{mathletters}
   \begin{equation}
      \bE = E(r,t) \,\hat\bz
   \end{equation}
   \begin{equation}
      \bH = H(r,t) \,\hat\bth
   \end{equation}
   \begin{equation}
      \bj_e = j_e(r,t) \,\hat\bz
   \end{equation}
   \begin{equation}
      \bj_g = j_g(r,t) \,\hat\bth
   \end{equation}
\end{mathletters}
Amp\`ere's Law (\ref{eq:ampere}) has only a $\hat\bz$ component,
\begin{equation}
   \frac{1}{r} \frac{\partial}{\partial r} \left(rH\right)
      - \frac{\partial E}{\partial t} = j_e
\>,
\label{eq:Az}
\end{equation}
and Faraday's Law (\ref{eq:faraday}) has only a $\hat\bth$ component,
\begin{equation}
   - \frac{\partial E}{\partial r} + \frac{\partial H}{\partial t} = -j_g
\>.
\label{eq:Fth}
\end{equation}
The two Gauss laws show that the charge densities are zero,
\begin{equation}
   \rho_e = \rho_g = 0
\>.
\end{equation}
Eqs.~(\ref{eq:Az}) and (\ref{eq:Fth}) are the equations of motion for
$E$ and $H$.

Near the ends of the flux tube, $|z|>z_1$, we must allow for new
components $E_r$ and $j_{er}$, as well as for $\rho_e$, and all fields
will be $z$-dependent.

%
\subsection{Vector potentials}

In order to calculate the time evolution of the matter fields, we need
the vector potentials.  In fact, we need only the dual potential
$B_\mu$ \cite{ref:onepot}, which enters the field equation for the
monopole field.  From (\ref{dual1}) and (\ref{dual2}),
\begin{eqnarray}
   \bE &=& -\nabla\times\bB +\hat\bz(\hat\bz\cdot\nabla)^{-1}\rho_e
   \label{eq:EB}
   \\
   \bH &=& -\nabla B^0 - \frac{\partial\bB}{\partial t}
      -(\hat\bz\cdot\nabla)^{-1}(\hat\bz\times\bj_e)
   \label{eq:HB}
\>,
\end{eqnarray}
where we have chosen $n^\mu=(0,\hat\bz)$.  We choose the gauge
$B_0=0$.  Inverting (\ref{eq:EB}) gives
\begin{equation}
   \bB = (\hat\bz\cdot\nabla)^{-1}(\hat\bz\times\bE)
\>.
\end{equation}
Our {\em ansatz\/} for $\bE$ then gives $\bB=B\,\hat\bth$, with
\begin{equation}
   B(r,z) = \int_{-\infty}^z E_r(r,z')\,dz'\>.
\label{BE}
\end{equation}
This integral only gets contributions from the regions around the
sources, since $E_r=0$ far from the charges.  This makes $B$
independent of $z$ in the central region $|z|<z_1$.  We can relate the
integral in (\ref{BE}) to the charge distribution near the sources as
follows.  We take a cylindrical surface of radius $r$ with ends at
$-\infty$ and $-z_1$.  Gauss' Law gives for this surface
\begin{equation}
   Q(r) = 2\pi r\int_{-\infty}^{-z_1} E_r(r,z) \,dz 
      + 2\pi \int_0^r E_z(r',-z_1)\,r'\,dr'
\>,
\end{equation}
where $Q(r)$ is the charge inside the cylinder (composed of the original
source plus the space charge around it). Noting that $E_z(r',-z_1)$ is
the $z$-independent $E_z(r')$, we obtain the relation
\begin{equation}
   B(r) = \frac{Q(r)}{2\pi r} - \frac1r\int_0^r E_z(r')\,r'\,dr'
\label{eq:BQE}
\end{equation}
and its inverse
\begin{equation}
   E \equiv E_z = - \frac1r\frac{\partial}{\partial r}(rB)
      + \frac1{2\pi r}\frac{\partial Q}{\partial r}
\>.
\label{eq:EBQ}
\end{equation}
Given $E$ and some model for the charge $Q(r)$ (see below),
we can calculate $B$.

\subsection{Magnetic monopoles}
The magnetic monopoles are represented by a classical Klein-Gordon
equation with a Higgs potential,
\begin{equation}
   D_\mu^B D^{\mu B}\psi + \lambda(|\psi|^2-v^2)\psi = 0
\>,
\end{equation}
where
\begin{equation}
   D^B_\mu \equiv \partial_\mu - igB_\mu
\>.
\end{equation}
We make the {\em ansatz\/} $\psi=\rho(r,t) e^{i\chi_g(\theta)}$, giving
\begin{equation}
   \left[\frac{\partial^2}{\partial t^2}
      -\frac1r\frac{\partial}{\partial r}r\frac{\partial}{\partial r}
      -\left(\frac1r\frac{\partial}{\partial\theta}-igB\right)^2\right]
      \rho e^{i\chi}+\lambda\left(\rho^2-v^2\right)\rho e^{i\chi}=0
\>.
\end{equation}
We assume $\chi_g=n\theta$ (for a flux tube with $n$ units of electric
flux), so
\begin{equation}
   \left[\frac{\partial^2}{\partial t^2}
      -\frac1r\frac{\partial}{\partial r}r\frac{\partial}{\partial r}
      +\left(\frac nr-gB\right)^2\right]
      \rho+\lambda\left(\rho^2-v^2\right)\rho = 0
\>.
\label{eq:KG}
\end{equation}
This is the equation of motion for $\rho$.  It feeds back into
Maxwell's equations via the magnetic current,
\begin{equation}
   \bj_g = 2g\rho^2\left(\nabla\chi_g - g\bB\right)
\end{equation}
which contains only a $\hat\bth$ component,
\begin{equation}
   j_g = 2g\rho^2\left(\frac nr - gB\right)\ .
\label{eq:jgth}
\end{equation}
As shown in the Appendix, $n$ represents electric flux coming in from
infinity, so that we will set $n=0$ for simplicity.  The flux
generated by $Q$ at the ends of the flux tube is {\em not} quantized.

\subsection{Charged matter}
We represent the electrically charged matter by classical two-fluid
magnetohydrodynamics.  The positively and negatively charged fluids
both have particle density $n_e(r,t)$ (so that $\rho_e=0$) and their
velocities are $\bv^{\pm}$.  The fluid motion is determined by Euler's
equations
\begin{equation}
   m \left[\frac{\partial \bv^{\pm}}{\partial t}
      + (\bv^{\pm} \cdot \nabla) \bv^{\pm}\right] = \pm e \bE \pm 
      e \bv^{\pm} \times \bH - \frac{1}{n_e}{\nabla P}
\end{equation}
and the continuity equation
\begin{equation}
   \frac{\partial n_e}{\partial t} 
      + \nabla \cdot (n_e \bv^{\pm}) = 0
\>.
\end{equation}
The electric current is given by
\begin{equation}
   \bj_e = n_e e \left( \bv^+ - \bv^- \right)
\>.
\end{equation}

We assume in the central region that all quantities depend only on $r$
and $t$, so that $\bv^{\pm}\cdot\nabla=v^{\pm}_r
\frac{\partial}{\partial r}$.  Recalling that $\bE=E\,\hat\bz$ and
$\bH=H\,\hat\bth$, we have
\begin{eqnarray}
   m\frac{\partial}{\partial t}v^{\pm}_r
      &=& -mv^{\pm}_r\frac{\partial}{\partial r}
      v^{\pm}_r \mp ev^{\pm}_zH - \frac{1}{n_e}\frac{\partial P}
      {\partial r}
   \nonumber \\
   m\frac{\partial}{\partial t}v^{\pm}_z
      &=& -mv^{\pm}_r\frac{\partial}{\partial r}
      v^{\pm}_z \pm eE\pm ev^{\pm}_rH
\end{eqnarray}
and
\begin{equation}
   \frac{\partial n_e}{\partial t} 
      = - v^{\pm}_r \frac{\partial n_e}{\partial r}
      - \frac{n_e}{r} \frac{\partial}{\partial r}(r v^{\pm}_r)
\>.
\end{equation}
Not surprisingly, these equations allow the {\em ansatz}
\begin{eqnarray*}
   v_r^+ &=& v_r^-\equiv v_r
   \\
   v_z^+ &=& -v_z^-\equiv v_z
\>,
\end{eqnarray*}
we find
\begin{eqnarray}
   m\frac{\partial}{\partial t}v_r &=& 
      -mv_r\frac{\partial}{\partial r}v_r -ev_zH - \frac{1}{n_e}
      \frac{\partial P}{\partial r}
   \nonumber \\
   m\frac{\partial}{\partial t}v_z &=& 
      -mv_r\frac{\partial}{\partial r}v_z +eE+ev_rH
   \nonumber \\
   \frac{\partial n_e}{\partial t} 
      &=& - v_r \frac{\partial n_e}{\partial r}
      - \frac{n_e}{r} \frac{\partial}{\partial r}(r v_r)
\>.
\label{eq:veom}
\end{eqnarray}
These equations feed back into Maxwell's equations via the current,
which is in the $\hat\bz$ direction and has the magnitude
\begin{equation}
   j_e = 2n_e e v_z
\>.
\label{eq:jez}
\end{equation}

We need an equation of state to relate $P$ to $n_e$.  Having in mind a
quark--gluon plasma, we choose the equation of state of a relativistic
ideal gas with the appropriate number of massless fermions and bosons.
We set the chemical potential to zero and write all quantities in
terms of the temperature,
\begin{eqnarray}
   P &=& 37 \frac{\pi^2}{90} T^4 \nonumber \\
   n_e &=& 37 \frac{\zeta(3)}{\pi^2} T^3
\>.
\label{eq:state}
\end{eqnarray}
Then eqs.~(\ref{eq:veom}) can be written in terms of the velocities and
temperature as
\begin{eqnarray}
   m\frac{\partial}{\partial t}v_r &=& 
      -mv_r\frac{\partial}{\partial r}v_r -ev_zH 
      - \frac{4 \pi^4}{90 \zeta(3)}
      \frac{\partial T}{\partial r}
   \nonumber \\
   m\frac{\partial}{\partial t}v_z &=& 
      -mv_r\frac{\partial}{\partial r}v_z +eE+ev_rH
   \nonumber \\
   \frac{\partial T}{\partial t} 
      &=& - v_r \frac{\partial T}{\partial r}
      - \frac{T}{3r} \frac{\partial}{\partial r}(r v_r)
\>.
\label{eq:vteom}
\end{eqnarray}

The advantage of a hydrodynamic description of the charged matter is that
the fluid interacts directly with the field strengths \bE\ and \bH.
If we were to consider classical or quantum field
theory instead of hydrodynamics, we would need the vector potential $\bA$. 
In our geometry, the simplest {\em ansatz\/} would require two components, 
$A_r$ and $A_z$. 
The potential would necessarily be $z$-dependent in the central region;
moreover,
the simplified treatment of the ends of the flux tube (see below) would
be impossible.

\subsection{The ends of the flux tube}

For study of the central region, all we need to know about the complex
regions at the ends of the flux tube is $Q(r)$, which represents the
charge density built up near $-z_0$ by the current $\bj_e$.  (The
region near $z=+z_0$ is, of course, a mirror image of the region near
$-z_0$.)  Without dealing in detail with the motion of charges near
the ends of the flux tube, we can guess at a few models:
\begin{enumerate}
\item {\em Point charge:\/} Here we just assume that all the current
  merely accumulates in a point charge at $z=-z_0$. Then $Q(r,t)$ is
  independent of $r$, and charge conservation gives
\begin{equation}
   \frac{dQ}{dt}=-2\pi\int_0^{\infty}j_e(r)\,r\,dr
\>.
\label{eq:qptchg}
\end{equation}
\item {\em Surface charge density:\/}
Here we let the charge pile up on a plate near $z=-z_0$, giving a surface
charge density $\sigma(r,t)$.  Then
\begin{equation}
   Q(r,t)=2\pi\int_0^r \sigma(r',t)\,r'\,dr'
\end{equation}
and
\begin{equation}
   \frac{\partial\sigma}{\partial t}=-j_e
\>.
\label{charge_cons}
\end{equation}
Initial conditions have to be specified for $\sigma$ to represent the
original source of the flux tube.
\item {\em Both:\/}
Keeping the initial charge pointlike, we put only the accumulated charge
into $\sigma$. Then
\begin{equation}
   Q(r,t) = Q_0+2\pi\int_0^r \sigma(r',t)\,r'\,dr'
\end{equation}
and $\sigma$ is determined by (\ref{charge_cons}) as above.  The initial
condition is $\sigma=0$.  Note that in this case the space charge will
never exactly cancel $Q_0$ and thus any plasma oscillations will be
asymmetric.
\end{enumerate}
The initial conditions must in any case satisfy Gauss' Law,
\begin{equation}
   2\pi\int_0^\infty E(r)r\, dr=Q(\infty)
\>,
\label{overallGauss}
\end{equation}
if there is no flux coming from infinity (i.e., $n=0$).

\subsection{Initial conditions}
We start off the system in the configuration of a static flux tube
with a stationary fluid in it.  We combine the static limit of
(\ref{eq:Fth}) with (\ref{eq:EBQ}) and (\ref{eq:jgth}) to give
\begin{equation}
   \frac{\partial}{\partial r} \frac{1}{r} \frac{\partial}{\partial r}
      \left[rB(r)-\frac1{2\pi}Q(r)\right]=2g^2\rho^2B 
\>,
\end{equation}
where we have chosen $n=0$.  Eq.~(\ref{eq:BQE}) gives the boundary
condition $B(0)=0$ unless $Q$ includes a point charge $Q_0$; in that
case
\begin{equation}
   B(r)\sim\frac{Q_0}{2\pi r}
\end{equation}
as $r\to0$.
Eq.~(\ref{eq:BQE}) together with (\ref{overallGauss}) gives
$rB(r)\to0$ at infinity.

The static limit of the Klein-Gordon equation (\ref{eq:KG}) is
\begin{equation}
   \left[-\frac1r\frac{\partial}{\partial r}
      r\frac{\partial}{\partial r} + g^2B^2\right]
      \rho +\lambda\left(\rho^2-v^2\right)\rho =0
\>,
\end{equation}
with the boundary conditions that $\rho$ is zero at the origin and
tends to $\rho=v$ at infinity.

Focusing on the case where $Q$ is a point charge, we define the
reduced field $b(r)$ via
\begin{equation}
   B=\frac Q{2\pi r}b
\>,
\end{equation}
giving
\begin{equation}
   b''-\frac{b'}r-2g^2\rho^2b=0
\>,
\label{eq:bstat}
\end{equation}
with $b(0)=1$ and $b(\infty)=0$, and
\begin{equation}
   -\rho''-\frac{\rho'}r
   +\left[\left(\frac{Qg}{2\pi}\frac br\right)^2
   +\lambda(\rho^2-v^2)\right]\rho=0
\>.
\label{eq:rhostat}
\end{equation}
These equations determine $\rho$ and $B$, and hence $E$, for the
static flux tube.  Clearly $H=0$ here.

\section{Plasma oscillations}
\label{sec:results}

We determine the time evolution of the system through the following system
of equations.
The Maxwell equations (\ref{eq:Az}) and (\ref{eq:Fth}) give \bE\ and
\bH; the MHD equations (\ref{eq:vteom}) give \bv\ and $T$, and hence
$\bj_e$ via (\ref{eq:jez}) and (\ref{eq:state}).
Eq.~(\ref{eq:qptchg}) gives the charge $Q$ whence (\ref{eq:BQE})
gives the vector potential \bB.
The scalar field $\rho$ evolves according to the Klein-Gordon equation
(\ref{eq:KG}), and (\ref{eq:jgth}) gives the magnetic current
$\bj_g$, the last ingredient for the Maxwell equations.
The initial conditions, as noted, consist of the static flux tube with
an initial value of $Q$, and of a static fluid distribution specified by
$T(r)$.

We choose four sets of parameters, listed in Table \ref{table:params}.
The magnetic charge $g$ and the vacuum expectation value
$v=\rho(r\to\infty)$ determine the vector mass $m_V=\sqrt2gv$ and its
reciprocal, the London penetration depth $\lambda_L=m_V^{-1}$.
The scalar self-coupling $\lambda$ determines, along with $v$,
the (dual) Higgs mass $m_H=\sqrt{2\lambda}v$ and the Ginzburg-Landau coherence 
length $\xi=m_H^{-1}$.
The ratio of these lengths is $\kappa = \lambda_L / \xi=\sqrt\lambda/g$. 
If this value is
smaller than one then the superconductor is of Type I; otherwise it is of
Type II.

In ordinary superconductors, flux tubes are observed only in Type II materials,
when the applied magnetic field lies between the critical values
$H_{c1}$ and $H_{c2}$ and penetrates via creation of an Abrikosov lattice.
Type I materials, on the other hand, expel the field entirely as long as
superconductivity persists.
The situation would be different if one were to introduce a pair of
magnetic monopoles.
Then the magnetic field would be forced into a flux tube between the
monopoles, even in a Type I material.
The difference between Type I and II would lie in the stability of the
flux tube against splitting into a lattice (a ``rope'') of smaller tubes:
Type I tubes would be stable, while Type II tubes would split.
This splitting is presumably inaccessible from our cylindrically symmetric
{\em ansatz}.

The values of $\lambda$ and $g$ fixed phenomenologically in \cite{ref:DGL}
are in the Type II region, and we choose these values for our Type II cases.
For Type I, we increase $g$ and decrease $\lambda$ to make $\kappa<1$.
We take $v=126$~MeV from \cite{ref:DGL} as well.
We fix the electric charge $e$ according to the Dirac quantization
condition $eg / 4\pi = 1$, even though this is not meaningful for
continuum hydrodynamics.

The charged fluid presents a conundrum.
Our intuition about confinement suggests that the superconducting vacuum
should expel this matter, and thus confine it to the flux tube.
(This happens in the Friedberg--Lee Model \cite{ref:Friedberg}.)
In this Abelian theory, however, the locally neutral fluid is {\em not\/}
confined.
If the initial conditions contain a fluid inside the flux tube only,
it will flow outward in a hydrodynamic rarefaction wave.
Perhaps the fluid outside the flux tube may be interpreted as a hadronic
fluid.
In any case, we prefer not to superimpose radial hydrodynamic flow on the
oscillations of the flux tube.
Thus we choose a {\em uniform\/} initial fluid density, outside the flux
tube as well as inside.
Having in mind a quark--gluon plasma,
we set the initial temperature to be
spatially uniform with $T_0$ = 150 MeV. 

The initial value $Q_0$ of the charge at the ends of the flux tube
gives the initial amount of electric flux. 
In a $pp$ collision, one would have $Q_0\sim e$, corresponding to the
creation of a flux tube in the fundamental representation;
an $AA$ collision between heavy nuclei would give\cite{ref:biro,ref:KMS}
$Q_0\sim A^{1/3}e$, which reaches $Q_0\sim6e$ for large nuclei.
A large value of $Q_0$ will give a thick flux tube and
large-amplitude plasma oscillations, and thus enhance the non-linear effects
due to the Higgs coupling.
Unfortunately, very large values of $Q_0$ are outside the range of stability of
our numerics; we choose values that are as large as possible given this
constraint.

The plasma frequency is
\begin{equation}
   \omega_p = \sqrt{\frac{2 n_e e^2}{m}}
\>,
\end{equation}
where the factor of two reflects the fact that we have two fluids.
For given $n_e$ and $e$, we fix $m$ so as to tune $\omega_p$ to either side
of the vector mass $m_V$; thus the plasma oscillations will occur
at frequencies either above or below
the threshold for radiation.
We expect {\em a priori} that $m$ will be in the neighborhood of the 
dynamical mass $eT$ of light particles in a heat bath.
The parameters used in
\cite{ref:DGL} give a rather small vector mass, and with $e$ = $4\pi / g$
the condition $\omega_p < m_V$
leads to an unreasonably large value of $m$. 
Instead we choose in this case
to lower the value of the electric charge to $e=1$.

\begin{figure}
   \begin{center} 
   \mbox{\epsfig{file=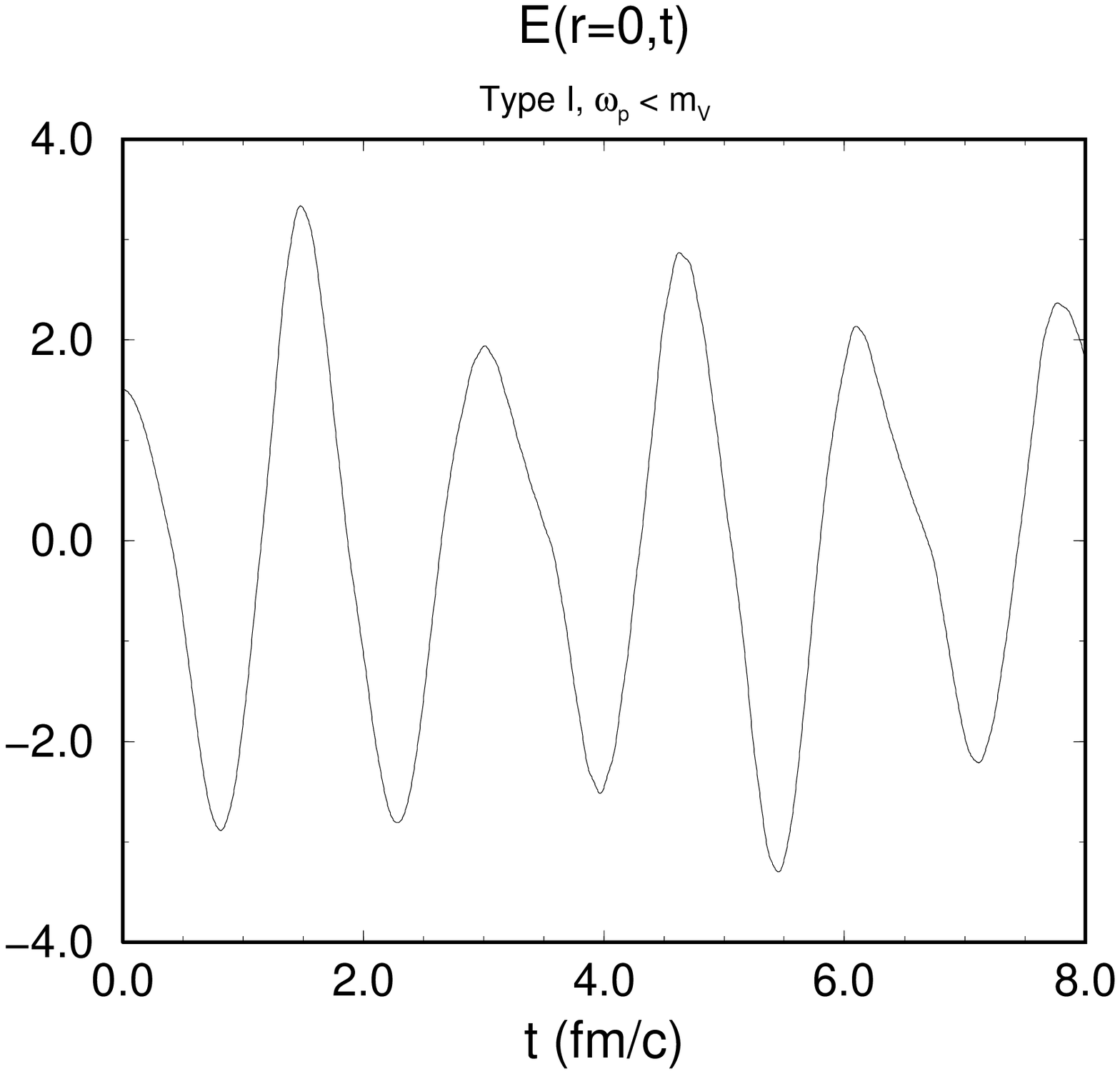,width=6.0in}}
   \caption{Electric field on the $z$ axis, for Type I superconductor with
    $\omega_p < m_V$.}
   \label{case1E}
   \end{center}
\end{figure}

\begin{figure}
   \begin{center} 
   \mbox{\epsfig{file=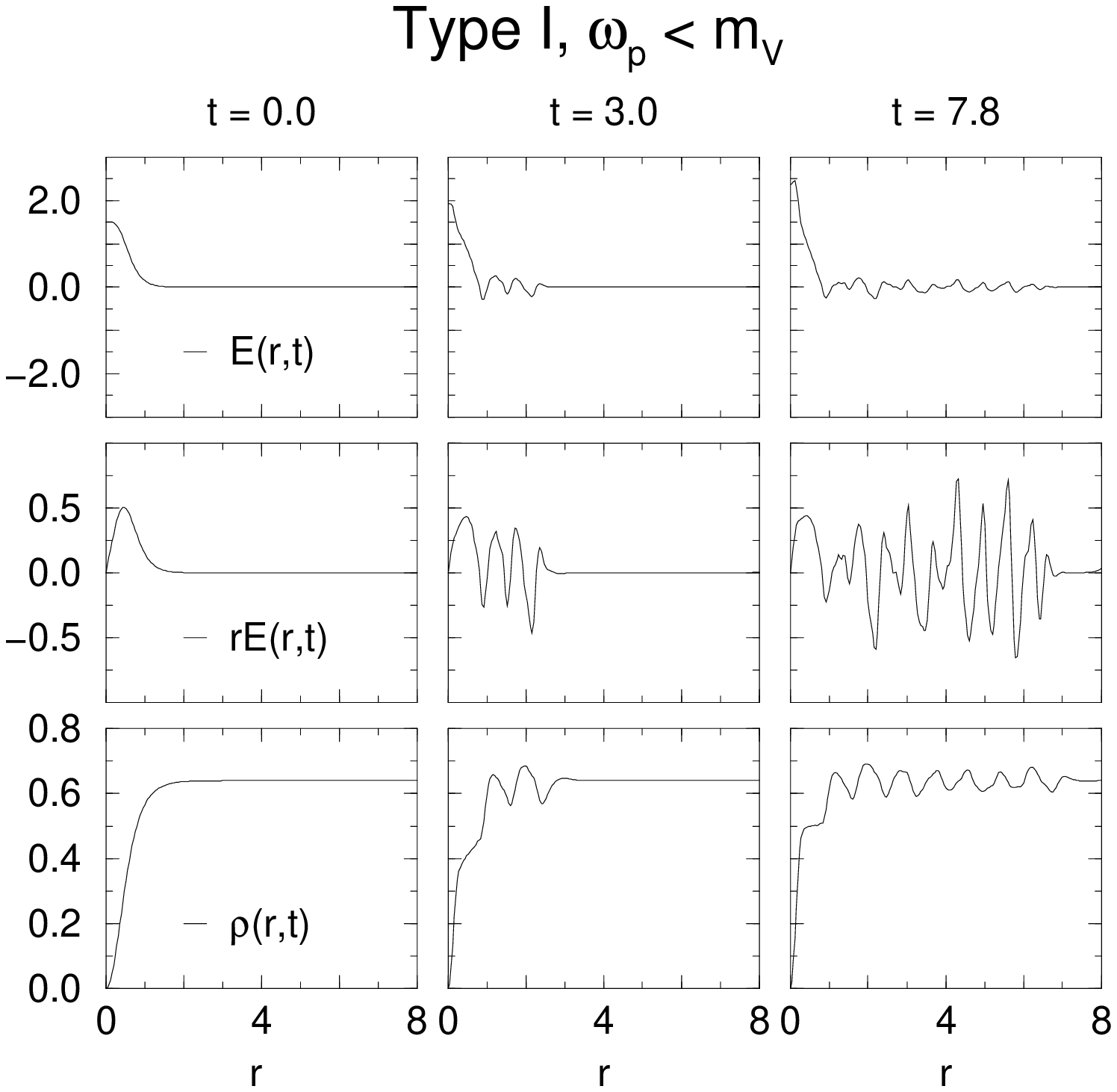,width=6.5in}}
   \caption{Snapshots of the electric field $E$ and monopole field $\rho$ at
   times of maxima in the on-axis field $E(r=0)$: Type I superconductor, 
   $\omega_p < m_V$.}
   \label{case1fields}
   \end{center}
\end{figure}

\paragraph*{Case 1.}
Fig.~\ref{case1E} shows plasma oscillations in the on-axis electric
field for the Type I superconductor with $\omega_p<m_V$.
The oscillations are clearly nonlinear, with varying amplitude and
misshapen waveform.
Fig.~\ref{case1fields} contains snapshots taken at the times of maxima in
$E(r=0)$.
These snapshots show radiation emanating from the flux tube in both
$E$ and $\rho$.
The plots of $rE$ show that the oscillating electric flux in the outgoing 
wave is as large in amplitude as the initial flux distribution.
It is difficult to tell, however, whether the wave amplitude decays
as $r^{-1/2}$, as expected for a true propagating wave, or as  $r^{-1}$,
which would make the wave evanescent.

\begin{figure}
   \begin{center} 
   \mbox{\epsfig{file=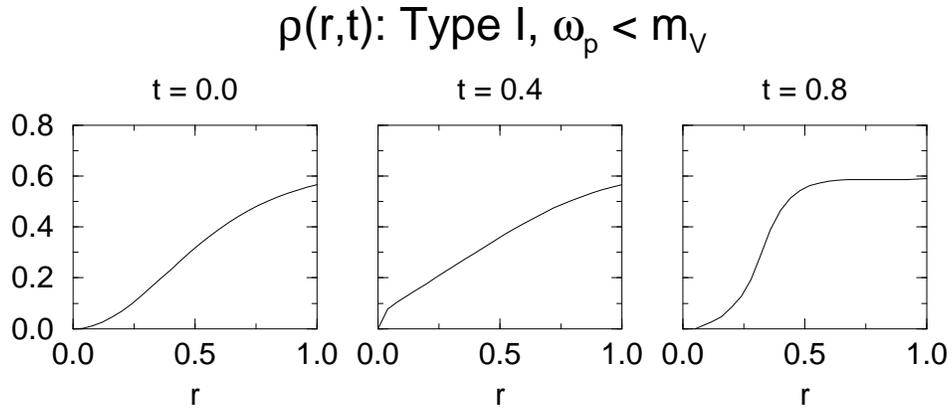,width=6.0in}}
   \vspace{-2.5in}
   \caption{Snapshots of
   $\rho(r,t)$ in the first oscillation: Type I superconductor, 
   $\omega_p < m_V$.}
   \label{case1rho}
   \end{center}
\end{figure}

Since the frequency of the plasma oscillations is below the vector mass,
this cannot be simple linear radiation.
The third row of fig.~\ref{case1fields}, indeed, shows strong coupling of
the nonlinear $\rho$ field to the electromagnetic wave.
The origin of this coupling, shown in fig.~\ref{case1rho}, is the
collapse of the flux tube under the pressure of the $\rho$ field when
the electric field is weak.
The figure shows snapshots of $\rho$ taken in the first half-cycle of
the evolution.
The middle snapshot, taken when $E(r=0)=0$, shows appreciable
narrowing of the flux tube core even as $\rho$ is driven downward at
larger radii in the first oscillation of the outgoing wave.

\begin{figure}
   \begin{center} 
   \mbox{\epsfig{file=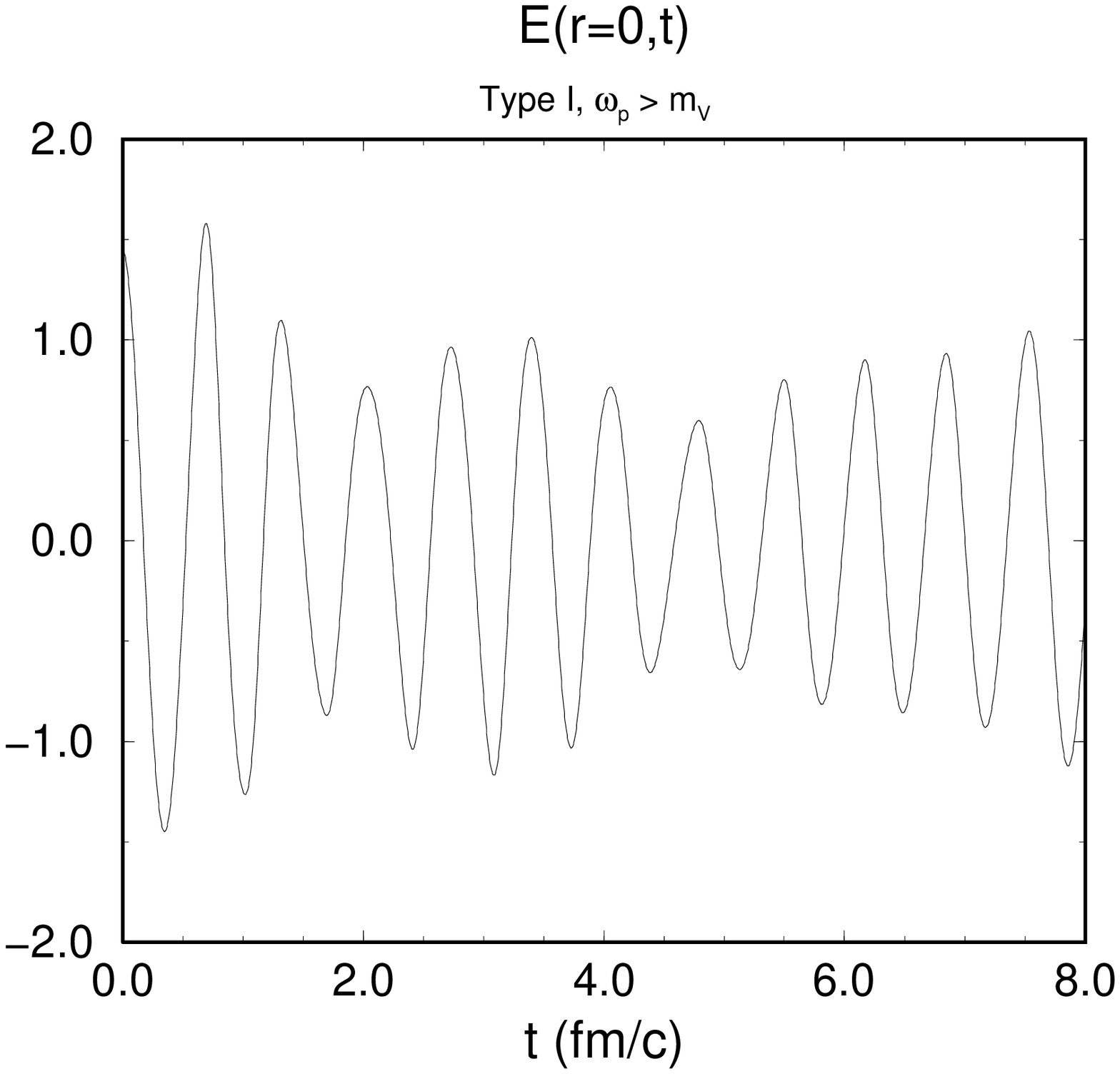,width=6.0in}}
   \caption{Electric field on the $z$ axis, for Type I superconductor with
   $\omega_p > m_V$.}
   \label{case2E}
   \end{center}
\end{figure}

\begin{figure}
   \begin{center} 
   \mbox{\epsfig{file=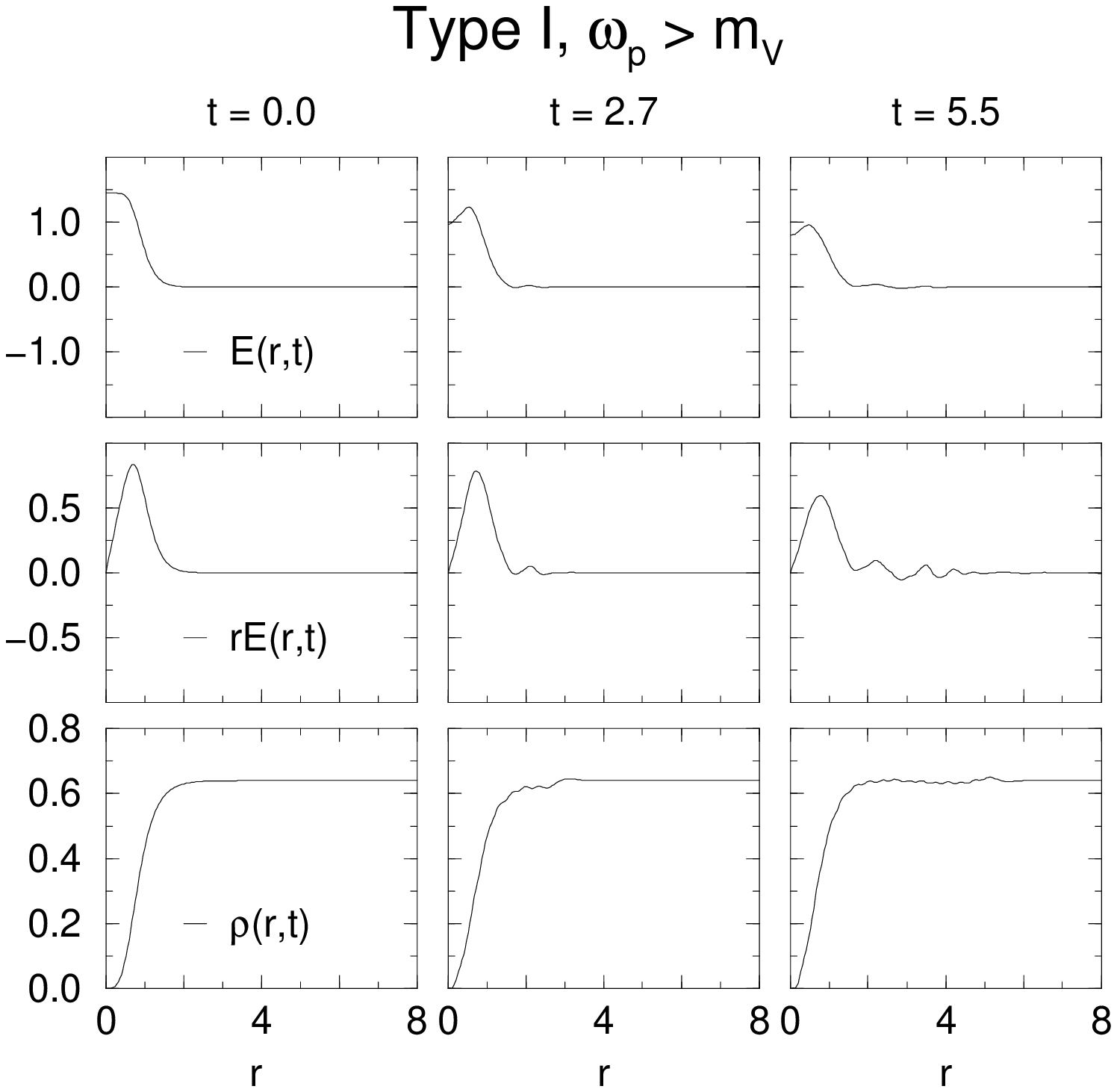,width=6.5in}}
   \caption{Snapshots of the electric field $E$ and monopole field $\rho$ at
   times of maxima in the on-axis field $E(r=0)$: Type I superconductor,
   $\omega_p > m_V$.}
   \label{case2fields}
   \end{center}
\end{figure}

\paragraph*{Case 2.}
Oscillations in the Type I superconductor with a larger plasma frequency,
$\omega_p>m_V$, are shown in figs.~\ref{case2E} and~\ref{case2fields}.
Here the irregularity of the waveform of $E(r=0)$ is due to ordinary
radiation in interaction with a more-or-less static flux tube wall.
Again there is appreciable electric flux radiated outward, but there is
little effect on the $\rho$ field.
This is because the latter cannot respond when driven by the high-frequency
electromagnetic oscillations.
We note that the larger value of $gQ_0$ here creates a thicker flux
tube and keeps more flux in the flux tube despite the radiation.

\begin{figure}
   \begin{center} 
   \mbox{\epsfig{file=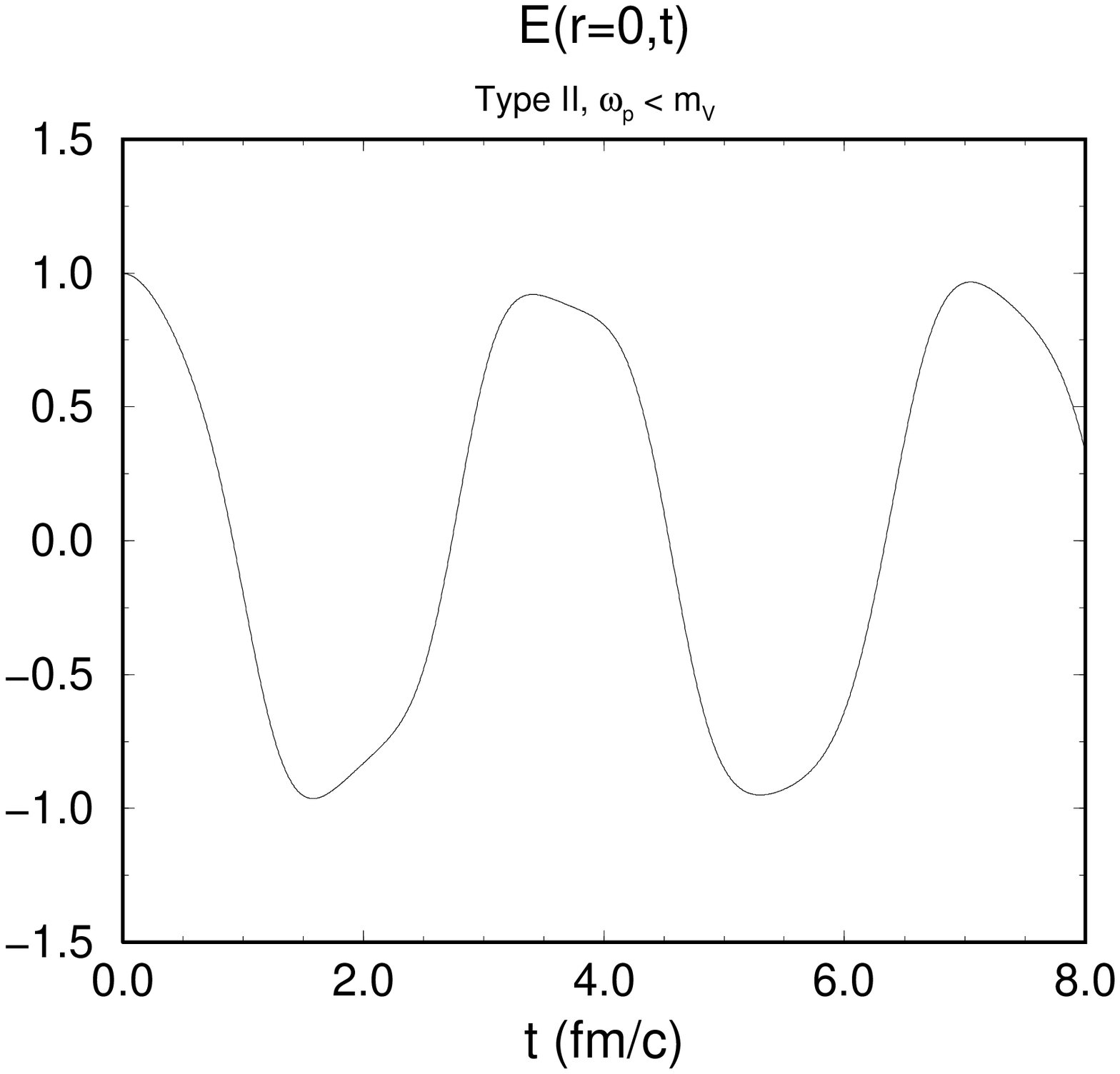,width=6.0in}}
   \caption{Electric field on the $z$ axis, for Type II superconductor with
   $\omega_p < m_V$.}
   \label{case3E}
   \end{center}
\end{figure}

\begin{figure}
   \begin{center} 
   \mbox{\epsfig{file=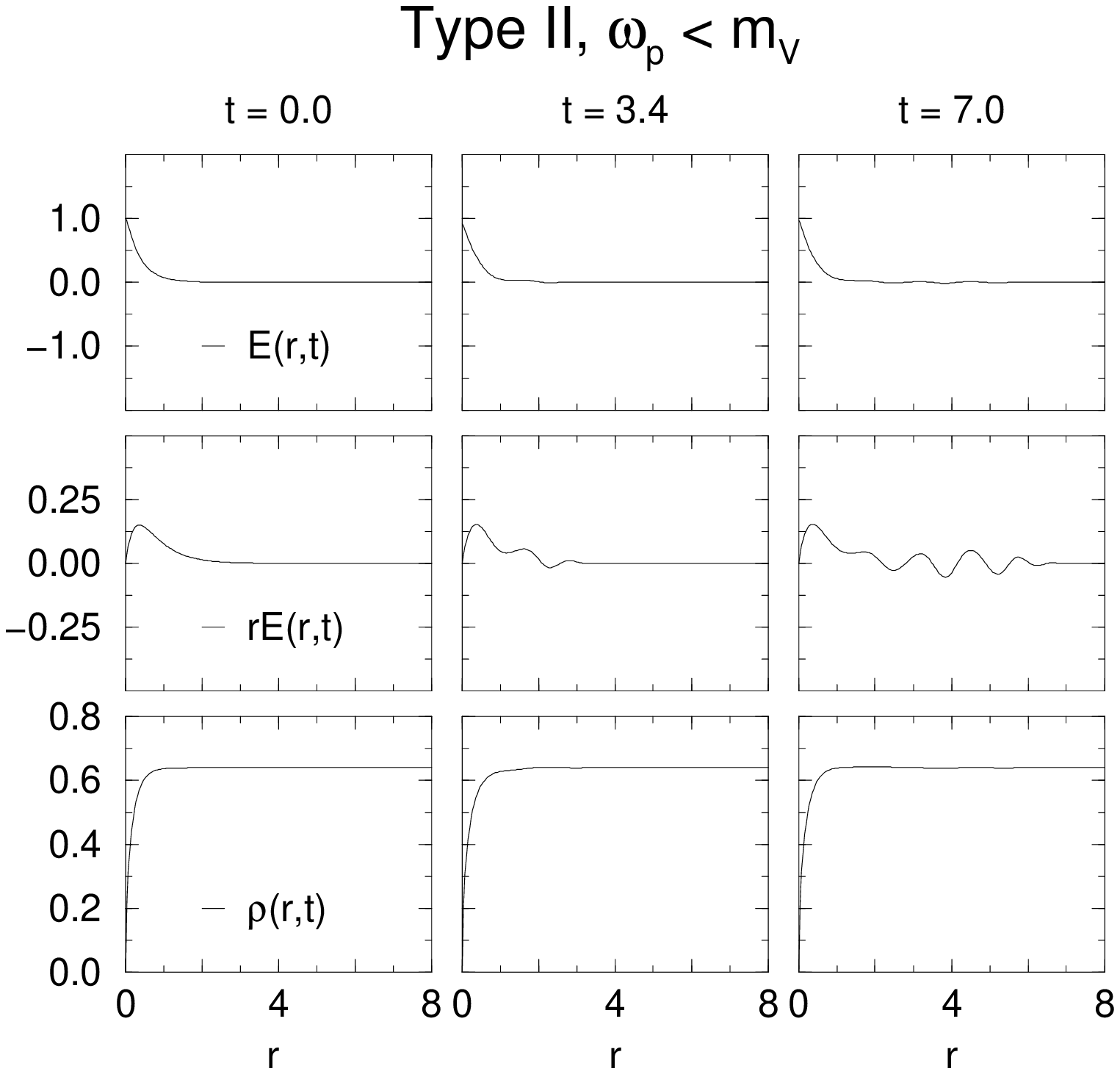,width=6.5in}}
   \caption{Snapshots of the electric field $E$ and monopole field $\rho$ at
   times of maxima in the on-axis field $E(r=0)$: Type II superconductor,
   $\omega_p < m_V$.}
   \label{case3fields}
   \end{center}
\end{figure}

\paragraph*{Case 3.}
The Type II superconductor with $\omega_p<m_V$ is similar in behavior to the 
Type I system.
The smaller plasma frequency makes it difficult to calculate
through more than two or three oscillations,
in spite of the high stability of the Crank-Nicholson algorithm we use.
In fig.~\ref{case3E} we see strong deformation of the waveform.
Fig.~\ref{case3fields} shows radiation similar to the Type I case, but 
the perturbation of the $\rho$ field is less apparent.

\begin{figure}
   \begin{center} 
   \mbox{\epsfig{file=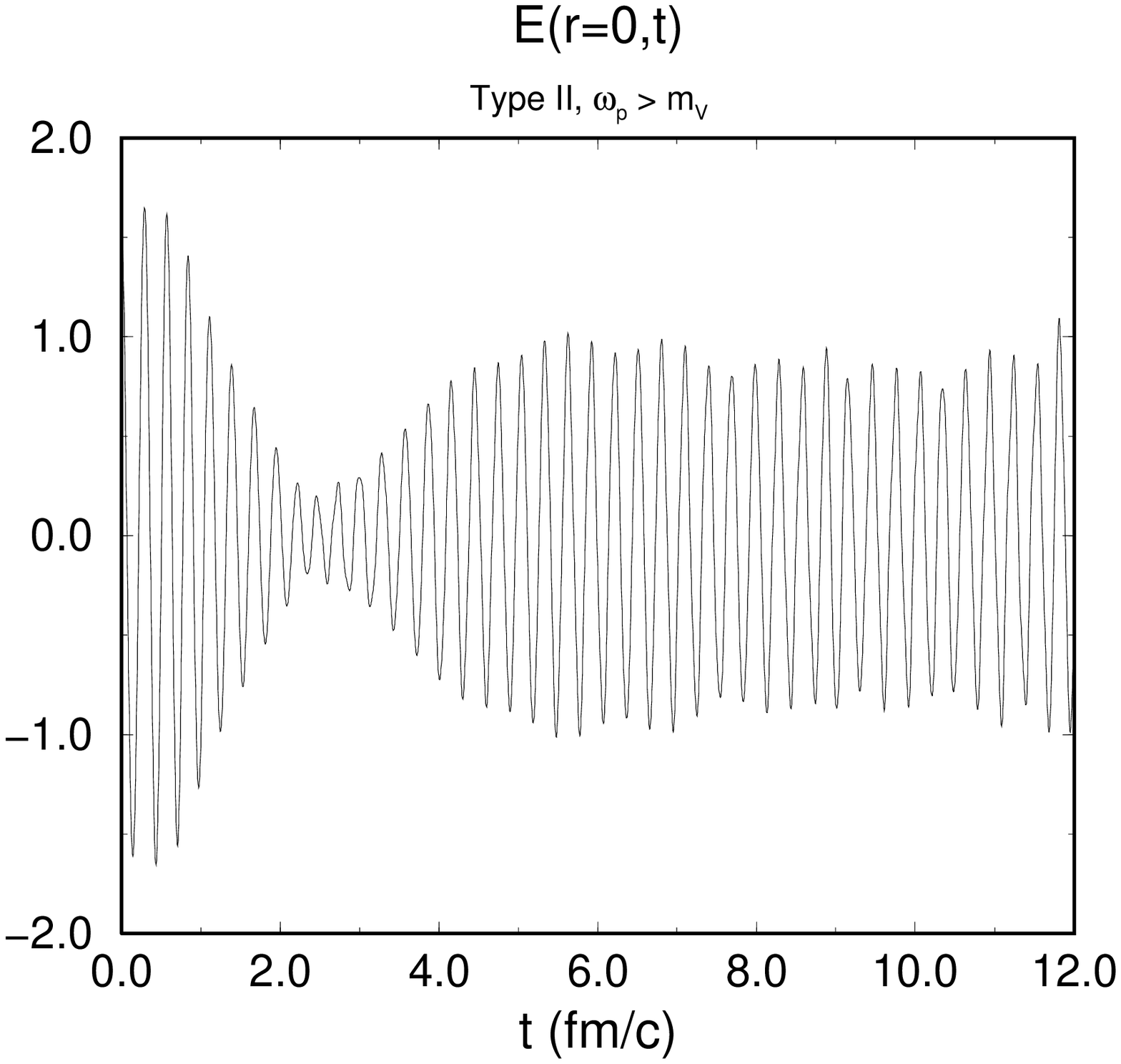,width=6.0in}}
   \caption{Electric field on the $z$ axis, for Type II superconductor with
   $\omega_p > m_V$.}
   \label{case4E}
   \end{center}
\end{figure}

\paragraph*{Case 4.}
Finally, the corresponding plots for the Type II superconductor with
$\omega_p>m_V$ show considerable structure.
Recall that this parameter set
is that used in \cite{ref:DGL} for
phenomenological fits to static quantities.\footnote{There is no electric
current in \cite{ref:DGL}, and hence no plasma frequency.}
The oscillations shown in fig.~\ref{case4E} are too rapid to couple
strongly to $\rho$, and we see strong electromagnetic radiation
(figs.~\ref{case4fields} and~\ref{case4fields_b}) accompanied by
weak high-momentum disturbances in $\rho$.

\section{Discussion}

Our numerical results raise difficulties for the dual superconductor
picture of confinement.
As a static model, the dual superconductor does indeed form flux tubes
that confine charges in string-like configurations.
Once the dynamics are examined, however, the lack of absolute color
confinement becomes apparent.
While the motion of neutral particle matter outside the flux tube
may be passed off as the emission of color-neutral hadrons, the
radiation of appreciable electric flux cannot.
The electric field in 't~Hooft's Abelian projection is after all
a {\em color\/} field, representing a coherent, colored gluon state.

It was predictable that oscillations with $\omega_p>m_V$ would
radiate into the Higgs vacuum, since the photon does have a less-than-infinite 
mass.
One might be tempted to restrict application of the dual superconductor model
to situations where frequencies are much less than $m_V$.
We have seen, however, that this is insufficient.
Even low-frequency plasma oscillations, where radiation should be
impossible, succeed in spreading out the electric flux via nonlinear
effects.
Thus the model is not reliable even in this regime.
It is possible that the model can be made applicable to low-frequency
physics by choosing a monopole Lagrangian more general than the
$\psi^4$ of the simplest Landau-Ginzburg theory.
Absolute confinement, however, will never be realized in this way.
(The introduction of a strongly nonlinear dielectric constant is what
enables the Friedberg--Lee model to confine all color fields,
but this model has not been derived from QCD.)

't Hooft's Abelian reduction rests on the identification of the
important degrees of freedom in an Abelian gauge.
It is supposed that the magnetic monopoles have a strong self-interaction,
leading to their condensation;
that the Abelian gluons, belonging to the Cartan subalgebra, turn this
condensate into a superconductor; and, most important, that the off-diagonal
gluons are irrelevant to confinement, except insofar as they screen
zero-triality states.
These off-diagonal gluons, however, retain all the self-couplings of the
original non-Abelian gauge theory.
We conjecture that they are essential to understanding time-dependent
phenomena related to confinement.\footnote{For recent work on deriving the
effective action of an Abelian reduction of QCD, see \cite{Kondo}.
Here the off-diagonal gluons are not neglected, but rather integrated
out explicitly.}
%
\section*{Acknowledgements}
We thank Dr.~Alex Kovner and Prof.~Eli Turkel for their assistance.
This work was supported by the Israel Science Foundation under Grant
No.~255/96-1 and by the Basic Research Fund of Tel Aviv University.

\appendix
\section*{Flux quantization}
Flux quantization, as it turns out, affects only flux coming in from
infinity and not the flux due to the charge $Q$ at the ends of the
flux tube.  The quantization condition comes from demanding that the
total energy be finite \cite{ref:coleman}.  The energy of the monopole
field contains the term
\begin{equation}
   E_{\text{static}}\equiv \int d^3x\,\left|{\bf D}^B \psi\right|^2
\>.
\end{equation}
As $x^2+y^2\to\infty$, we have $\psi\to\psi_0e^{i\chi(\theta)}$ and 
$\bB\to B\,\hat\bth$, and thus the integrand approaches
\begin{equation}
   \varepsilon_\infty\equiv \psi_0^2\cdot\left(
     \frac{\chi'(\theta)}r-gB\right)^2
\>.
\end{equation}
In order that $\int^\infty r\,dr\,d\theta\,\varepsilon$ be finite, we must
have at large $r$ that
\begin{equation}
   \frac{d\chi}{d\theta}=gBr
\>.
\label{circle}
\end{equation}
Since $\chi$ is only determined up to a multiple of $2\pi$, it can gain
such a multiple when one goes around the circle.
Integrating (\ref{circle}) over the circle at fixed $r$ gives
\begin{equation}
   2\pi n= g\int Br\,d\theta=g\oint\bB\cdot d{\bf l}
\>,
\end{equation}
where $n$ is an integer.  In view of (\ref{eq:EB}),
\begin{equation}
   \oint\bB\cdot d{\bf l}=\int(\nabla\times\bB)\cdot d{\bf S}
      =-\int \bE\cdot d{\bf S} + Q=-\Phi_E+Q
\end{equation}
where $\Phi_E$ is the total electric flux.
Thus
\begin{equation}
   \Phi_E=Q-2\pi n/g
\>.
\label{quantE}
\end{equation}
Only the {\em external\/} flux, the flux that does not end at $Q$, is
quantized in units of $2\pi/g$.  Choosing $\Phi_E=Q$, i.e., no
external flux, means setting $n=0$.



\begin{table}
\caption{Parameters used in numerical simulations. $v=126$~MeV in all cases,
and the initial temperature is $T_0=150$ MeV.}
\begin{tabular}{ldddd}
   {} & Type I & Type I & Type II & Type II\\
   {} & $\omega_p < m_V$ &  $\omega_p > m_V$ 
   & $\omega_p < m_V$ & $\omega_p > m_V$ \\ \hline
   $\lambda$ & 10. & 10. & 25. & 25. \\
   $g$ & 5.5 & 5.5 & 2.3 & 2.3 \\
   $e$ & 2.3 & 2.3 & 1.0 & 5.5 \\
   $m_V$ (MeV) & 975. & 975. & 411. & 411. \\
   $m_H$ (MeV) & 565. & 565. & 893. & 893. \\
   $\lambda_L$ (fm) & 0.20 & 0.20 & 0.48 & 0.48 \\
   $\xi$ (fm) & 0.35 & 0.35 & 0.22 & 0.22 \\
   $\kappa$ & 0.58 & 0.58 & 2.18 & 2.18 \\
   $m$ (MeV) & 250. & 50. & 250. & 50. \\
   $\omega_p$ (MeV) & 790. & 1766. & 343. & 4192. \\ 
   $T_p=2\pi/\omega_p$ (fm) & 1.57 & 0.70 & 3.60 & 0.30 \\
   $Q_0$ & 2.3 & 4.6 & 1.0 & 4.6 \\
\end{tabular}
\label{table:params}
\end{table}


\begin{figure}
   \begin{center} 
   \mbox{\epsfig{file=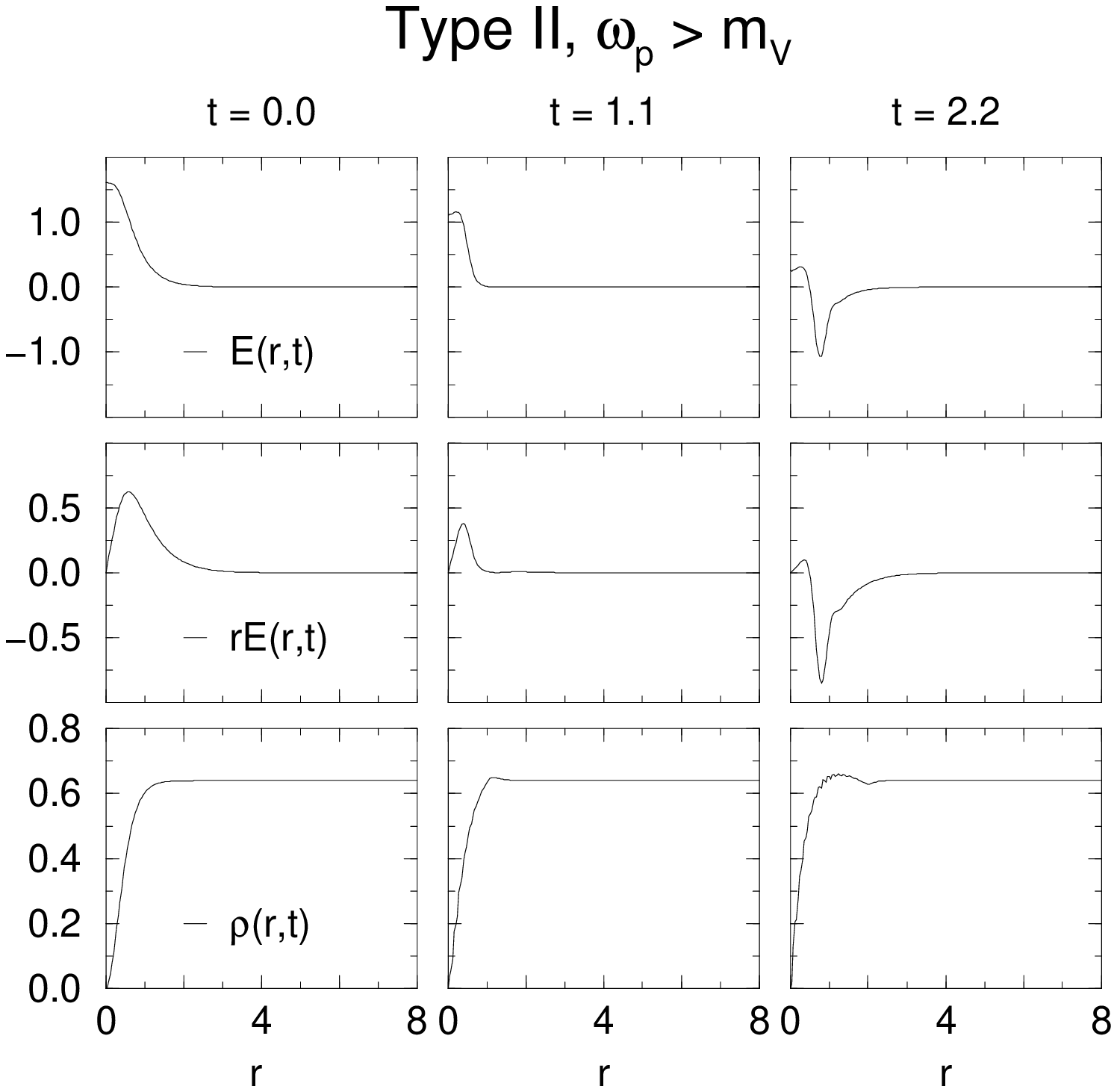,width=6.5in}}
   \caption{Snapshots of the electric field $E$ and monopole field $\rho$ at
   times of maxima in the on-axis field $E(r=0)$: Type II superconductor,
   $\omega_p > m_V$.}
   \label{case4fields}
   \end{center}
\end{figure}

\begin{figure}
   \begin{center} 
   \mbox{\epsfig{file=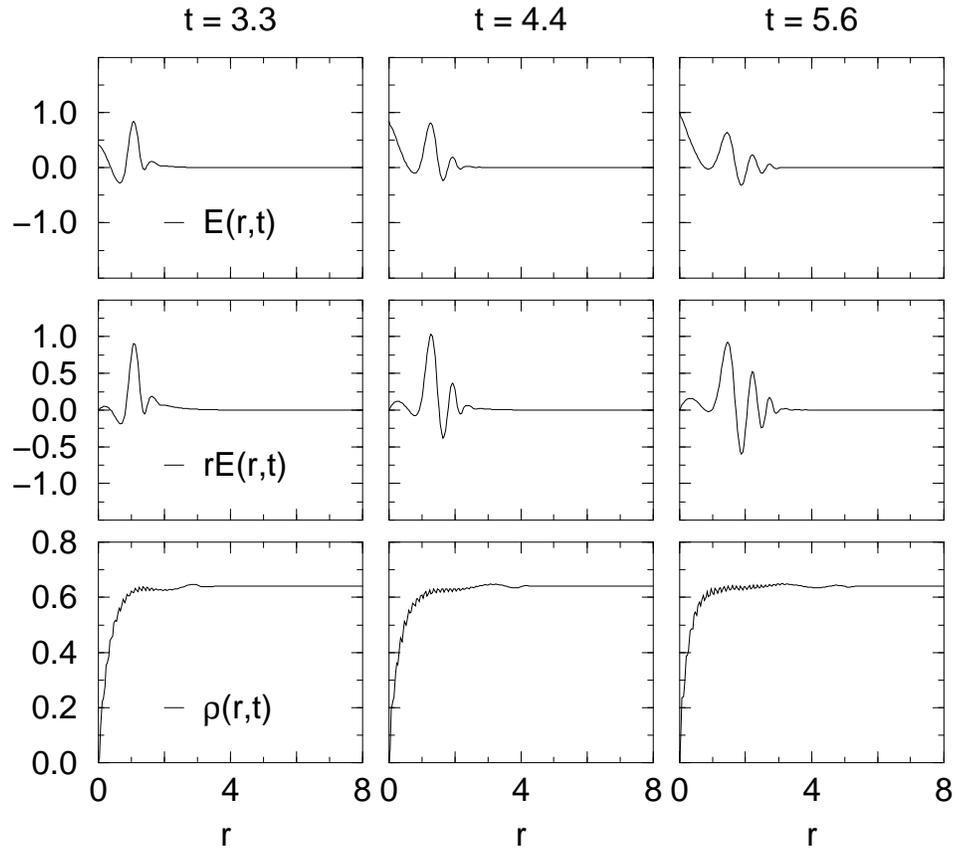,width=6.5in}}
   \caption{Continuation of fig.~\ref{case4fields}.}
   \label{case4fields_b}
   \end{center}
\end{figure}

\end{document}